\documentclass{aa} 
\usepackage{graphicx}
\usepackage{txfonts}
\def\sn{\hbox{S/N}}  
  
\def\vsin{\hbox{$v \sin i$}}  
  
\def\kms{\hbox{km\,s$^{-1}$}}

\def\degr{\hbox{$^\circ$}}

\def\kis{\hbox{$\chi^2$}}   
\def\kisr{\hbox{$\chi^2_{\rm r}$}}

\begin{document}
   \title{The rapid rotation and complex magnetic field geometry of Vega\thanks{Based on observations obtained at the Bernard Lyot Telescope (TBL, Pic du Midi, France) of the Midi-Pyr\'en\'ees Observatory, which is operated by the Institut National des Sciences de l'Univers of the Centre National de la Recherche Scientifique of France, and at the Canada-France-Hawaii Telescope (CFHT) which is operated by the National Research Council of Canada, the Institut National des Sciences de lÕUnivers of the Centre National de la Recherche Scientifique of France, and the University of Hawaii.}}

   \author{P. Petit
          \inst{1}
          \and
          F. Ligni\`eres
          \inst{1}
          \and
          G.A. Wade
          \inst{2}  
          \and  
          M. Auri\`ere
          \inst{1}
          \and  
          T. B\"ohm
          \inst{1}
          \and
          S. Bagnulo
          \inst{3}
          \and  
          B. Dintrans
          \inst{1} 
          \and  
          A. Fumel
          \inst{1} 
          \and  
          J. Grunhut
          \inst{2} 
          \and  
          J. Lanoux
          \inst{4} 
          \and
          A. Morgenthaler
          \inst{1} 
          \and  
          V. Van Grootel
          \inst{1} 
          }

   \offprints{P. Petit}

   \institute{
   Laboratoire d'Astrophysique de Toulouse-Tarbes, Universit\'e de Toulouse, CNRS, France \\ 
   \email{petit@ast.obs-mip.fr, ligniere@ast.obs-mip.fr, auriere@ast.obs-mip.fr,\\ boehm@ast.obs-mip.fr, amorgent@ast.obs-mip.fr, dintrans@ast.obs-mip.fr, aurelie.fumel@ast.obs-mip.fr, vvangroo@ast.obs-mip.fr}
               \and
             Department of Physics, Royal Military College of Canada, PO Box 17000, Station Forces, Kingston, Ontario, Canada \\ 
             \email{Gregg.Wade@rmc.ca, Jason.Grunhut@rmc.ca}
            \and
             Armagh Observatory, College Hill, Armagh BT61 9DG, Northern Ireland, U.K.\\
             \email{sba@arm.ac.uk}
            \and
             Centre d'\'etude Spatiale des Rayonnements, Universit\'e de Toulouse, CNRS, France \\ 
             \email{joseph.lanoux@cesr.fr}
             }

   \date{Received ??; accepted ??}

 
  \abstract
   {The recent discovery of a weak surface magnetic field on the normal intermediate-mass star Vega raises the question of the origin of this magnetism in a class of stars that was not previously known to host detectable magnetic fields.}
   {We aim to confirm the field detection reported by Ligni\`eres et al. (2009) and provide additional observational constraints about the field characteristics, by modelling the large-scale magnetic geometry of the star and by investigating a possible seasonal variability of the reconstructed field topology.}
   {We analyse a total of 799 high-resolution circularly-polarized spectra collected with the NARVAL and ESPaDOnS spectropolarimeters during 2008 and 2009. Using about 1,100 spectral lines, we employ a cross-correlation procedure to compute, from each spectrum, a mean polarized line profile with a signal-to-noise ratio of about 20,000. The technique of Zeeman-Doppler Imaging is then used to determine the rotation period of the star and reconstruct the large-scale magnetic geometry of Vega at two different epochs.}
   {We confirm the detection of circularly polarized signatures in the mean line profiles. The signal shows up in four independent data sets acquired with both NARVAL and ESPaDOnS. The amplitude of the polarized signatures is larger when spectral lines of higher magnetic sensitivity are selected for the analysis, as expected for a signal of magnetic origin. The short-term evolution of polarized signatures is consistent with a rotational period of $0.732 \pm 0.008$~d. The reconstruction of the magnetic topology unveils a magnetic region of radial field orientation, closely concentrated around the rotation pole. This polar feature is accompanied by a small number of magnetic patches at lower latitudes. No significant variability in the field structure is observed over a time span of one year.}
   {The repeated observational evidence that Vega possesses a weak photospheric magnetic field strongly suggests that a previously unknown type of magnetic stars exists in the intermediate-mass domain. Vega may well be the first confirmed member of a much larger, as yet unexplored, class of weakly-magnetic stars now investigatable with the current generation of stellar spectropolarimeters.}


   \maketitle

\section{Introduction}

In a well-established picture, two distinct magnetic regimes exist in main-sequence stars of intermediate mass, with Ap/Bp stars hosting strong, structured magnetic fields (e.g. Wade et al. 2000, Auri\`ere et al. 2007) while fields with large-scale components can be excluded above a level of a few gauss in other types of stars belonging to the same mass domain (e.g. Shorlin et al. 2002, Wade et al. 2006, Auri\`ere et al. 2007, 2010). This simple picture was recently shaken by the detection of a weak magnetic field on the normal A star Vega (Ligni\`eres et al. 2009, hereafter L09), suggesting that a potentially significant fraction of A stars might display a similar type of magnetism that has so far escaped the scrutiny of observers because it was hidden under the detection threshold of stellar spectropolarimetric observations. In the face of such an exciting finding, the next step is then to confirm the field detection of L09 and gather further observational clues to help understand the origin of this previously unknown type of magnetic field. 

Two main scenarios are usually proposed to account for the origin of stellar magnetic fields. The first option attributes a fossil nature to the magnetic field, in the sense that the field is inherited from star formation or an early convective evolutionary phase, the strength of which has been amplified during stellar contraction (e.g. Moss 2001). This first model is generally preferred to account for the strong magnetism of chemically-peculiar stars, since it can be reconciled with the observed simplicity of their field geometries (dominated by a dipole), their diverse geometrical characteristics and the absence of correlation between stellar rotation and field strength (e.g. Hubrig et al. 2000).  A second option invokes the continuous generation of the magnetic field through dynamo processes active either within the convective core (Brun et al. 2005) or in the radiative layers (e.g. Ligni\`eres et al. 1996, Spruit 2002). 

To help identify a theoretical framework applicable for Vega, it is crucial to accumulate more information about its surface magnetic structure and about the temporal evolution of the photospheric magnetic distribution. We propose here to progress in this direction, using high signal-to-noise ratio (\sn\ hereafter) spectropolarimetric observations that we model using the Zeeman-Doppler Imaging technique (Donati \& Brown 1997). 

We divide our study as follows : (a) we describe the high-resolution, high \sn\ spectropolarimetric time-series obtained for Vega, from which we confirm the detection of a photospheric magnetic field; (b) we investigate the rotational modulation of Zeeman signatures and reconstruct a magnetic map of Vega for two observing epochs; (c) we compare the vector magnetic distribution of the two maps and report the absence of any significant evolution of the field structure within one year; (d) finally, we discuss this new observational information and summarize our main results.

\section{Results}
\label{sect:results}

\subsection{Observing material}

\begin{table*}
\begin{center}
\caption[]{Journal of observations.}
\begin{tabular}{cccccc}
\hline
Instrument & Date & no. spectra & subexp. time & Stokes V duration & Average \sn \\
                    &           &                      &         (sec)      &        (sec)                 & (LSD) \\
\hline
NARVAL & 25 Jul. 08 & 33 & 6 & 145&  $20750 \pm 3062$ \\
  & 26 Jul. 08 & 97 & 6 & 145 & $21705 \pm 865$ \\
  & 27 Jul. 08 & 97 & 6 & 145 & $21549 \pm 1943$ \\
  & 28 Jul. 08 & 30 &  6 & 145 & $17225 \pm 1829$ \\
\hline
NARVAL & 22 Jun. 09 & 5 & 16 & 185 & $16913 \pm 874$ \\
 & 23 Jun. 09 & 6 & 16 & 185 & $29562 \pm 1256$ \\
 & 27 Jun. 09 & 5 & 16 &185 &  $20776 \pm 934$ \\
 & 04 Jul. 09 & 33 & 16 & 185 & $25596 \pm 6128$ \\
 & 05 Jul. 09 & 31 & 16 & 185 & $26964 \pm 3683$ \\
\hline
ESPaDOnS & 08 Sep. 09 & 99 & 4 & 135 & $22190 \pm 1428$\\
 & 09 Sep. 09 & 82 & 4 & 135 & $22438 \pm 1011 $\\
 & 10 Sep. 09 & 135 & 4 & 135 & $19304 \pm 3099 $\\
\hline
NARVAL & 26 Oct. 09 & 35 & 12 & 170 & $16829 \pm 730$ \\
 & 27 Oct. 09 & 51 & 12 & 170 & $19855 \pm 946$ \\
 &  31 Oct. 09 & 58 & 12 & 170 & $23561 \pm 4674$ \\
 &  01 Nov. 09 & 1 & 12 &170 &17598 \\
 &  03 Nov. 09 & 1 & 12 & 170 &15990\\
\hline
\end{tabular}
\end{center}
\noindent Notes : from left to right, we list the instrument used for observing, the date, the number of spectra collected during the night, the exposure time of individual subexposures, the total duration of a Stokes V sequence (including detector readout and rotation of Fresnel rhombs) and the average and standard deviation of the \sn\ values of Stokes V LSD profiles across the night (calculated for spectral bins of 1.8 \kms).
\label{tab:obs}
\end{table*}

The observational material employed in this study consists of high-resolution spectra obtained simultaneously in classical spectroscopy (Stokes I) and circularly polarized light (Stokes V) using the stellar spectropolarimeters NARVAL (T\'elescope Bernard Lyot, Observatoire du Pic du Midi, France) and ESPaDOnS (Canada-France-Hawaii Telescope, Hawaii). These two twin instruments provide full coverage of the optical spectral domain (370 nm to 1,000 nm) in a single exposure, at a resolving power of 65,000, with a peak efficiency of about 15\% (telescope and detector included). They consist of a bench-mounted spectrograph (based on a dual-pupil optical design and stored in a double thermal enclosure for optimal wavelength stability), fibre-fed from a Cassegrain-mounted module where the polarimetric analysis is performed prior to any oblique reflection of the beam. A series of 3 Fresnel rhombs (two half-wave rhombs that can rotate about the optical axis and one quarter-wave rhomb) are employed to perform a very achromatic polarimetric analysis over the whole spectral domain. They are followed by a Wollaston prism which splits the incident light into two beams, respectively containing light linearly polarized perpendicular/parallel to the axis of the prism. The two beams produced by the Wollaston prism are imaged onto the two optical fibres that carry the light to the spectrograph. Each Stokes V spectrum is obtained from a combination of four sub-exposures taken with the half-wave rhombs oriented at different azimuths (Semel et al. 1993). The data reduction is performed by Libre-Esprit, a dedicated, fully automated software described by Donati et al. (1997) and implementing the optimal spectral extraction principle of Horne (1986) and Marsh (1989).

A total of 838 spectra was recorded from July 2008 to October 2009, during 4 different telescope campaigns. The first data set, taken with NARVAL during 4 consecutive nights in July 2008, is described by L09. To complement this first time-series, we have recorded another set of 80 NARVAL spectra in June/July 2009, followed by 316 ESPaDOnS spectra in September 2009 and 146 NARVAL spectra in October/November 2009. The observations of July 2008 and September 2009 dominate all other available data due to their dense temporal sampling over consecutive nights, by their high and homogeneous signal-to-noise ratio and by the large number of spectra collected at these two epochs (about 70\% of the observing material at our disposal). For this reason, these two data sets will be preferentially used in the rest of our analysis. To ensure an optimal data quality in our study, we have discarded from our data sets all spectra in which a low \sn\ reveals significant atmospheric absorption or tracking problems. We have also ignored all spectra with significant solar contamination in the Stokes I parameter (observations collected close to sunrise). After cleaning up the data sets, we end up with a total of useful 799 spectra (Table \ref{tab:obs}). The integration time adopted for the four individual subexposures constituting the Stokes V sequences is varying from one observing run to the next, with values ranging from 4 sec in September 2009 to 16 sec in June 2009. To this shutter time, we must add another 120 sec for each Stokes V sequence, including readout time and rotation of the polarimetric optics. The total time spent to obtain a Stokes V spectrum is therefore comprised between 136 sec and 184 sec, depending on the observing run. 

The Least-Squares Deconvolution technique (LSD; Donati et al. 1997) was applied to all spectra, extracting from each of the 799 spectra a mean line profile with enhanced \sn. Different line lists were employed to check the robustness of our results, but unless specifically mentioned hereafter, the line list used in this paper is identical to that presented by L09. The mask is based on a Kurucz atmospheric model with an effective temperature T$_{\rm eff} = 10,000$~K, a surface gravity $\log(g)=4.0$ and a solar metallicity, yielding a total of about 1,100 atomic lines in the spectral window of NARVAL and ESPaDOnS. The resulting \sn\ of Stokes V LSD cross-correlation profiles is listed in Table \ref{tab:obs}.   

\subsection{Axi-symmetric magnetic component}

\begin{figure}
\centering
\includegraphics[height=7.5cm]{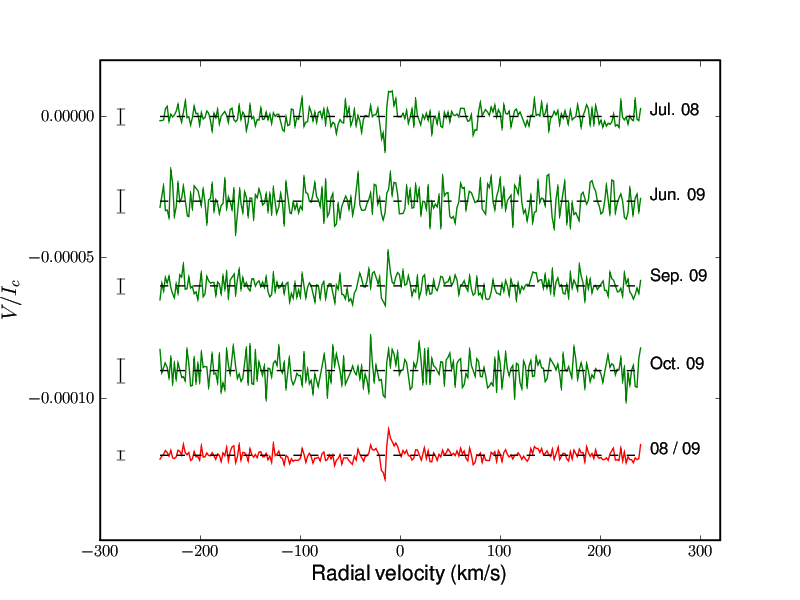}
\includegraphics[height=7.5cm]{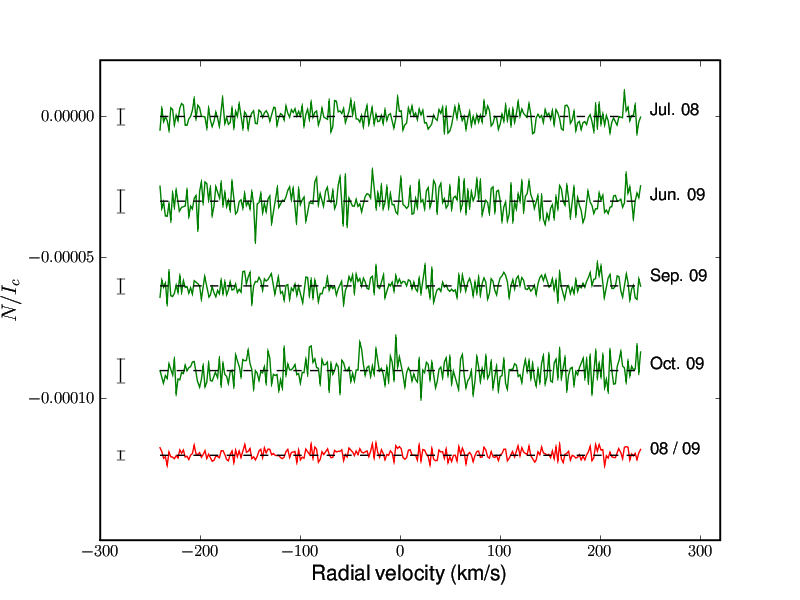}
\caption{Top panel : Averaged Stokes V LSD profiles of Vega for the various observing runs (green lines). The red line is obtained by averaging all 799 Stokes V profiles at our disposal. Note that successive profiles are shifted vertically for better clarity. The error bar corresponding to each profile is plotted at left. Bottom panel : same as upper part of the plot, for the "null" control profiles.}
\label{fig:stokesv}
\end{figure}

We do not detect any Zeeman signature in any of the individual Stokes V LSD profiles, in agreement with the previous analysis of L09. As a strategy to further improve the \sn\ of our data and lower our detection threshold, we repeat the approach of L09 and calculate a weighted average of all LSD profiles obtained in each observing run. We therefore end up with 4 distinct averaged Stokes V profiles, plus grand average obtained by averaging all LSD profiles collected in 2008 and 2009 (Fig. \ref{fig:stokesv}, upper frame).

All averaged profiles show a polarized signal located around the central radial velocity of the intensity profile (about -13.7~\kms), although the signal detection is ambiguous for the two observing runs suffering from the highest noise levels. Using a \kis\ test similar to the one proposed by L09, the false-alarm probability of a detection is equal to $3\times 10^{-11}$ in July 2008,  $3\times 10^{-1}$ in June 2009, $7\times 10^{-3}$ in September 2009, $10^{-1}$ in October 2009 and $10^{-15}$ in the grand average. Compared to the first averaged profile obtained in 2008, no statistically significant difference is observed in the signatures obtained during the 2009 observing runs. We emphasize that a similar signature is obtained using either NARVAL or ESPaDOnS data, and with or without inserting the Atmospheric Dispersion Corrector (ADC) in the beam prior to the polarimeter (observations from 2008 were taken without the ADC). The grand average LSD profile, obtained by grouping all available profiles together, has a noise level of $1.6\times 10^{-6}I_c$ and displays a polarized signal antisymmetric about the line centre, with a peak-to-peak amplitude of  $1.8\times 10^{-5}I_c$. The full velocity width of the signature is about 20~\kms (or 10 velocity bins), showing that it is comfortably resolved by our instrumental setup.

By using another combination of the four sub-exposures constituting the Stokes V spectrum, it is possible to calculate a "null" line profile which should contain no stellar polarized signal and from which many spurious instrumental signatures can be diagnosed. Similarly to L09, we do not detect any significant spurious signature when running this control calculation (Fig. \ref{fig:stokesv}, lower panel), suggesting that most instrumental effects are kept below a limit of about  $10^{-6}I_c$. Any significant spurious signature generated by a variability in the shape of Stokes I profiles (for instance, owing to the presence of stellar pulsations) can be ruled out by this test, since (a) its associated signal should show up in the control profile as well and (b) the line variability is unlikely to be confined to the line-center only, contrary to the observed Stokes V signal. A detailed investigation of Stokes I variability, based on the same observing material, is presented by B\"ohm et al. 2010 (in prep). 

\begin{figure}
\centering
\includegraphics[height=7.5cm]{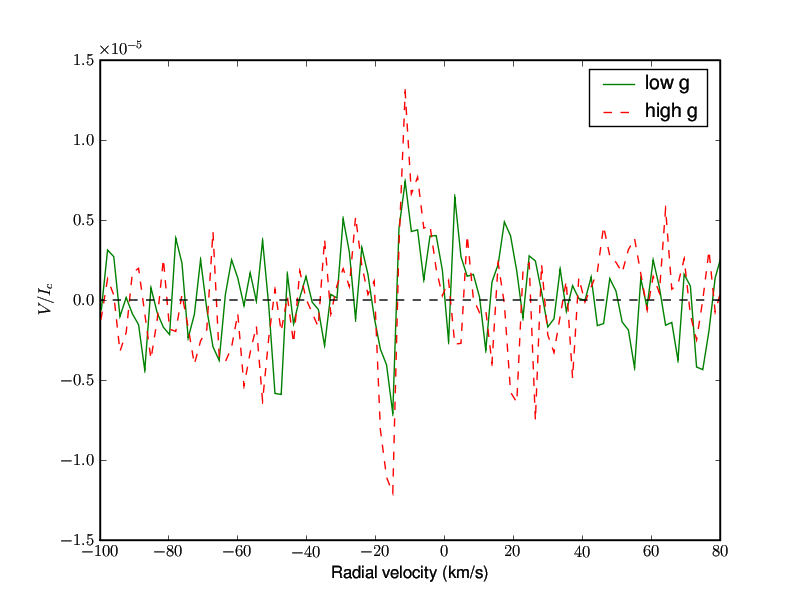}
\caption{Averaged Stokes V LSD profiles of Vega for different values of the Land\'e factor. The profile with a low Land\'e factor is plotted in green/solid line. The red/dashed line represents the effect of a line list with high Land\'e factors.}
\label{fig:lande}
\end{figure}

To further evaluate the proposed magnetic nature of the polarized signature, we calculate again the averaged Stokes V profile for the global data set, but this time we generate two distinct line-lists from our initial one, by defining a threshold in the Land\'e factor. The first sub-list contains all lines with a Land\'e factor $g \le 1.2$, which yields a mean Land\'e factor $\bar{g} = 0.94$. The second one is built up from lines with Land\'e factors $g \ge 1.2$, yielding $\bar{g} = 1.51$. We note that the average wavelength of both line-lists used for comparison are quite similar (495~nm for the low $\bar{g}$, against 498~nm for the higher $\bar{g}$). The resulting Stokes V profiles are plotted in Fig. \ref{fig:lande}, after correcting the Stokes V parameters for the difference in line depth of about 10\% observed in Stokes I. Signatures with similar shapes are derived from both line-lists, with a higher noise than previously observed while taking all available spectral lines in the analysis, because of the smaller number of spectral features used in the cross-correlation process. A difference in amplitude is visible between the two profiles, with a ratio of 1.72 between the peak-to-peak amplitudes obtained with the two masks (note that this difference cannot be taken as statistically significant, given the noise level). With a factor of 1.6 between the $\bar{g}$ values of the masks, the observed amplitude ratio is close to the ratio of land\'e factors, providing us with another hint in favour of the magnetic nature of the polarized signal. 

As previously proposed by L09, we also compare averaged profiles obtained from the red and blue halves of the line-list alone (not pictured here), reaching the conclusion that red and blue profiles exhibit a similar signature. Here we propose to check further the robustness of the field detection by employing yet another line-list, obtained by cleaning up our initial line-list, in which the mismatch between the theoretical atmospheric model and the actual spectra can possibly reduce our detection accuracy. First, we remove from the analysis all atomic lines suffering from significant blending with broad lines (e.g. Balmer lines) or molecular bands. By doing so, the number of lines drops to about 600. We then adjust the line weights of the model to agree better with the observed weights and compute again LSD profiles with the new mask. In spite of a lower number of lines in the analysis, the outcome of this test is to enhance the fit of the observed spectrum, thanks to the better match between the modelled and observed spectral lines. The outcome is especially spectacular in Stokes I, where the rms deviation of the fit to the observed spectrum is divided by a factor of $\approx 2$. In spite of the smaller number of spectral features used for cross-correlation, the more accurate line definition results in a marginally increased \sn\ of the Stokes V LSD profiles. The resulting Zeeman signature (not shown here) is consistent with that obtained using our initial mask. 

Finally, we use the centre-of-gravity technique (Rees \& Semel 1979) to derive an estimate of the longitudinal magnetic field from the Stokes I and V LSD profiles. The longitudinal field (expressed in gauss) is obtained through the equation~:

 \begin{equation}
B_l = -2.14\times10^{11}\frac{\int vV(v)dv}{\lambda_0 gc\int(I_c - I(v))dv}
\end{equation}

\noindent where $v$ (km\,s$^{-1}$) is the radial velocity, $\lambda_0$ (nm) the mean wavelength of the line-list used to compute the LSD profiles (496~nm here), $g$ the mean Land\'e factor (equal to 1.20) and $c$ (km\,s$^{-1}$) the light velocity. Choosing also a $\pm 30$~\kms\ radial velocity range around the line center as integral boundaries, we get $B_l = 0.6 \pm 0.2$~G while considering all data together, in agreement with L09, except for a slightly sharper error bar. By running this estimate again after replacing the Stokes V line profile by the null profile, we obtain $B_l = 0.2 \pm 0.2$~G.

\subsection{Zeeman-Doppler Imaging}
\label{sect:procedure}

To model the full time-series of Stokes V line profiles (without the previous restriction of considering phase-averaged profiles only), we make a model of a synthetic, spherical stellar surface divided into a grid of pixels, each pixel being associated with a local Stokes I and V profile. Assuming a given magnetic field strength and orientation for each pixel, local Stokes V profiles are calculated under the weak-field assumption, where Stokes V is proportional to $g.\lambda_0^2 .B_\parallel .\partial I / \partial \lambda$. $\lambda_0$ is the average wavelength of the LSD profile (about 496 nm for Vega), $B_\parallel$ is the line-of-sight projection of the local magnetic field vector, $g$ is the effective land\'e factor of the LSD profile (equal to 1.2) and $\partial I / \partial \lambda$ the wavelength derivative of the local synthetic Stokes I line profile (assumed to possess a Gaussian shape). We further assume that there are no large-scale brightness inhomogeneities over the stellar surface, so that all synthetic Stokes I profiles are locally the same over the whole visible photosphere. This last assumption is clearly wrong for Vega, where fast rotation is generating a significant gravity darkening (Takeda et al. 2008). However, several studies have already concluded that the content of magnetic maps issued from Zeeman-Doppler Imaging is mostly insensitive to the value of the input limb-darkening parameter (e.g. Petit et al. 2008), so that, by roughly behaving in a similar manner, the gravity darkening is also not a critical parameter since its influence is dominated by the projection effect that reduces the contribution of pixels located close to the stellar limb. A series of tests, using various approximate laws to describe the gravity/limb darkening, confirmed that this parameter does not affect significantly the results presented here. Finally, we chose to weight the amplitude of the local Stokes profiles for pixels located on the visible hemisphere according to a linear limb-darkening coefficient equal to 0.5 (Claret 2003). The central wavelength of each local profile is shifted according to the line-of-sight velocities of individual pixels, assuming $v.\sin i$ = 22 \kms\  and $i=7$\degr\ (Takeda et al. 2008). We note that the low inclinaison angle implies that low-latitude regions are always seen close to the limb, so that the combined effect of geometrical projection, limb darkening and gravity darkening tend to limit their contribution to the line profiles and affect the reconstruction accuracy of low-latitude magnetic features.

Synthetic Stokes V profiles are computed for all observed rotation phases and compared to the observations. The model adjustment is iterative and based on a maximum entropy algorithm (Skilling \& Bryan 1984). The version of the code used here makes a projection of the surface magnetic field onto a spherical harmonics frame (Donati et al. 2006), with the magnetic field geometry resolved into its poloidal and toroidal component (Chandrasekhar 1961). We limit the spherical harmonics expansion to $\ell<10$, since no improvement in the fit to the data is achieved by increasing further the maximum allowed value for $\ell$. 

\subsection{Rotation period}
\label{sect:period}

\begin{figure}
\centering
\includegraphics[width=10cm]{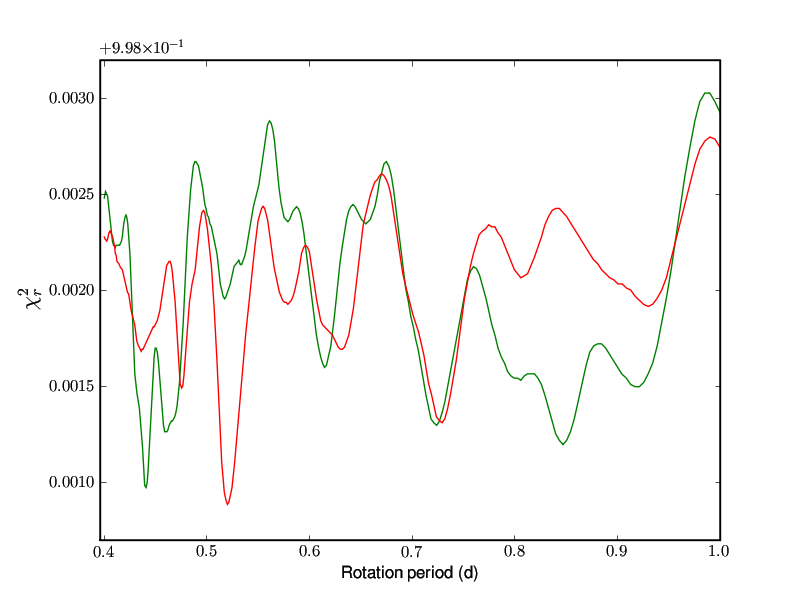}
\caption{Periodograms obtained from the NARVAL data set of July 2008 (green line) and from the ESPaDOnS observations of September 2009 (red curve). The numerical value $9.98\times 10^{-1}$ indicated on top left must be added to the values of the vertical axis.}
\label{fig:period}
\end{figure}

\begin{figure}
\centering
\includegraphics[width=9cm]{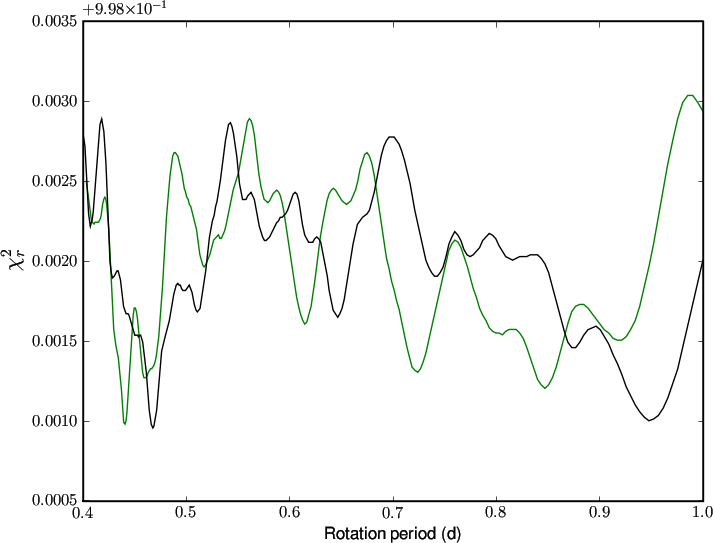}
\includegraphics[width=9cm]{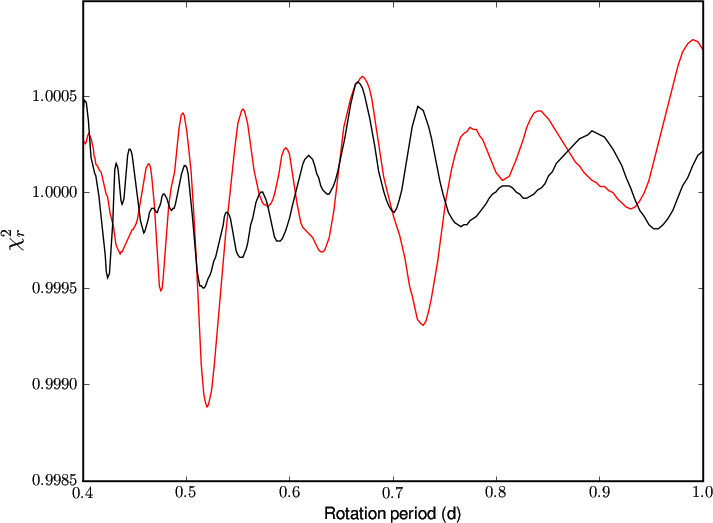}
\includegraphics[width=9cm]{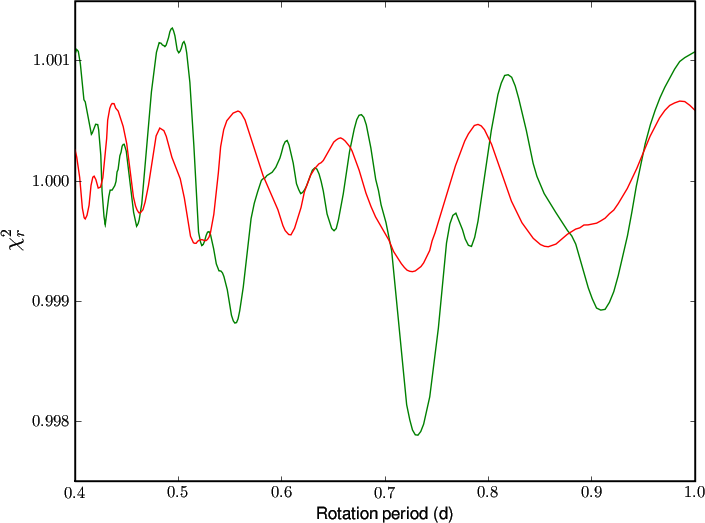}
\caption{Top panel : same as Fig. \ref{fig:period}, but this time the periodogram for July 2008 (green line) is compared to a shuffled data set, in which each individual Stokes V profile has been randomly associated to the Julian date of another profile of the same time-series (black line). Middle panel : same as the upper part of the plot, but with observations of September 2009 (red line) and their shuffled counterpart (black line). Bottom panel : same as Fig. \ref{fig:period}, but using two sets of fake Stokes V profiles computed by ZDI using our time-sampling of July 2008 and September 2009 (green and red lines, respectively).}
\label{fig:shuffle_jul08}
\end{figure}

The first step to reconstruct a relevant topology of the surface field consists in determining the stellar rotation period. To do so, we follow the approach of Petit et al. (2002) where a set of magnetic maps is calculated, assuming for each map a different value for the rotation period. We impose a constant entropy for all images and calculate a reduced \kis\ (\kisr\ hereafter) by comparing the set of synthetic Stokes V profiles produced by ZDI to the observed time-series of profiles. The resulting \kisr\ variations (plotted in Fig. \ref{fig:period}) are recorded over the range of rotation periods to determine the period value producing the best magnetic model (identified by the lowest value of the reduced $\chi^2$ goodness-of-fit parameter). Here, we scan 300 values of the period between 0.4~d and 1~d, a range that encompasses the various rotation periods already proposed in the literature (Aufdenberg et al. 2006, Takeda et al. 2008, Hill et al. 2010). Apart from fluctuations observed across the period span, the average \kisr\ value of our magnetic model is slightly different for the July 2008 and September 2009 data sets, with mean values of 0.89 and 0.94, respectively. To ease the comparison between both epochs, the resulting \kisr\ plots illustrated in Fig. \ref{fig:period} are compensated for this difference, so that the average of each curve on display is set to unity. 

Several minima are observed in the periodograms obtained for each data set. In July 2008, the best fit is reached at a period of about 0.44~d. In September 2009, the best modelling is obtained for a period close to 0.52~d. Among the observed minima, only one is consistently recovered in both data sets, corresponding to a rotation period close to 0.7~d. This common \kisr\ minimum is the second deepest in September 2009 and the fourth in July 2008. From the common minimum, we derive a rotation period of $0.725\pm 0.008$~d in July 2008, against $0.729\pm 0.008$~d in September 2009 (error bars are calculated according to Press et al. 1992). A similar period search, conducted after grouping all available data from 2008 and 2009, is strongly affected by the sparse observing window and remains unconclusive by exhibiting a dense forest of \kisr\ minima (not pictured here) over the same period range, with no preferred value of the period.

To estimate the possible impact that the time sampling adopted to acquire the data may have on the locations of the observed \kisr\ minima, we run again the same period search, but this time we shuffle the spectra by attributing to each spectrum the Julian date of another spectrum from the same time series, with a random date permutation. The outcome of this test is illustrated in Fig. \ref{fig:shuffle_jul08}.   Using the shuffled time-series, we obtain again a series of \kisr\ minima over the period domain. The observed \kisr\ variations display a significant correlation between the shuffled and actual time series (with a correlation coefficient of about 0.3, translating into a false-alarm probability of correlation close to $10^{-8}$), suggesting an actual impact of the observing window in our period search. The Pearson correlation coefficient obtained by comparing the actual periodograms is higher and equal to 0.39 (for a false-alarm probability of correlation of $10^{-12}$). The \kisr\ minima recovered using the proper time-series are generally lower than those derived from the shuffled ones, showing that a better magnetic model is obtained using properly ordered observations (a good hint that usable rotationally-modulated signal is present in our data). No \kisr\ minimum is showing up around 0.7~d in the shuffled periodograms, suggesting that this feature is not an artifact generated by the time sampling.

Finally, to test further the robustness of our approach, we replace the observed polarized profiles by synthetic profiles computed by ZDI in the reconstruction of the two magnetic maps described below in Sect. \ref{sect:maps}. These two sets of fake profiles are computed assuming a same rotation period of  0.7319~d (see below) and for the same time-sampling as the observed data. We add a Gaussian noise to the new profiles, in order to simulate a \sn\ of 20,000 (in rough agreement with our observations). The periodograms derived from the fake observations are plotted in the lower panel of Fig. \ref{fig:shuffle_jul08}. The \kisr\ variations are, in this case again, organized in a series of minima, with a Pearson coefficient of 0.3 between the two data sets and a false-alarm probability of correlation equal to $10^{-8}$. The \kisr\ minimum corresponding to the input period is showing up in both time-series, and this time this common minimum is the deepest at both epochs.

In the rest of the analysis, we adopt a value of 0.7319~d for the rotation period. This value is inside the error bars derived above and provides us with the best match between the reconstructed magnetic topology of the star at the two epochs (see below). We choose the Julian date $JD = 2454101.5$ (2007 January 01 at 00h00 UT) as phase origin.

\subsection{Magnetic topology}
\label{sect:maps}

\begin{figure}
\centering
\includegraphics[width=6cm]{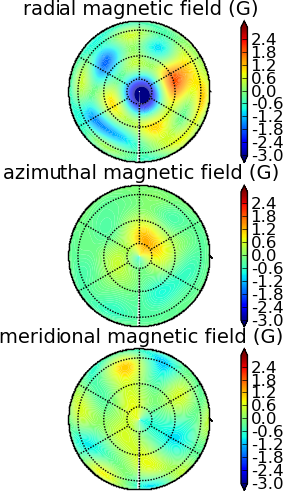}
\caption{Vectorial magnetic map of Vega for 2008 July, in polar projection. The 3 charts illustrate the field projection onto one axis of the spherical coordinate frame with the radial, azimuthal, and meridional field components. The magnetic field strength is expressed in gauss. The phase origin is set at the bottom of each chart and rotational phases are increasing in the clockwise direction.}
\label{fig:mapjul08}
\end{figure}

\begin{figure}
\centering
\includegraphics[width=6cm]{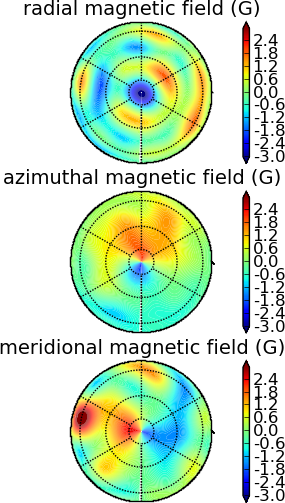}
\caption{Same as Fig. \ref{fig:mapjul08} for September 2009.}
\label{fig:mapsep09}
\end{figure}

\begin{figure}
\centering
\includegraphics[height=7.5cm]{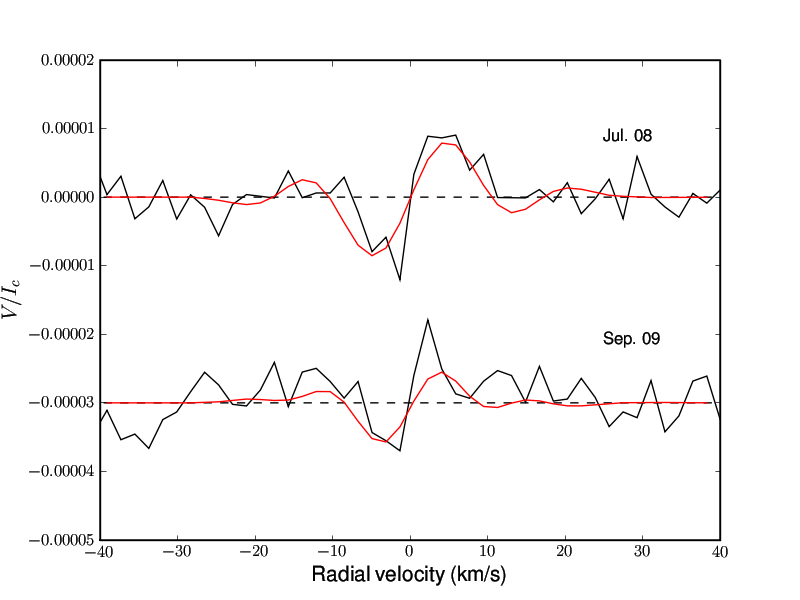}
\caption{Comparison between the observed averaged Stokes V LSD profiles and the synthetic averaged profiles of our magnetic model. The observations are illustrated in black (after correction for the average radial velocity of the star), while synthetic profiles are displayed in red.}
\label{fig:zdi_comp}
\end{figure}

\begin{figure}
\centering
\includegraphics[width=9cm]{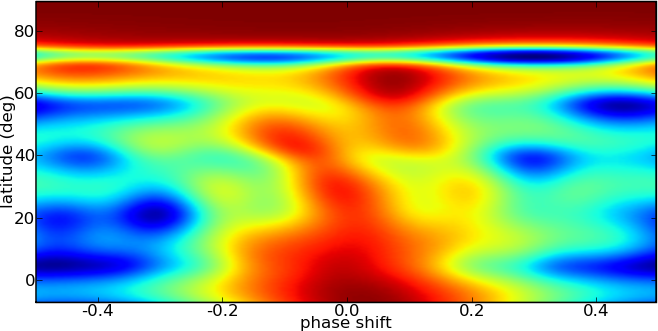}
\includegraphics[width=9cm]{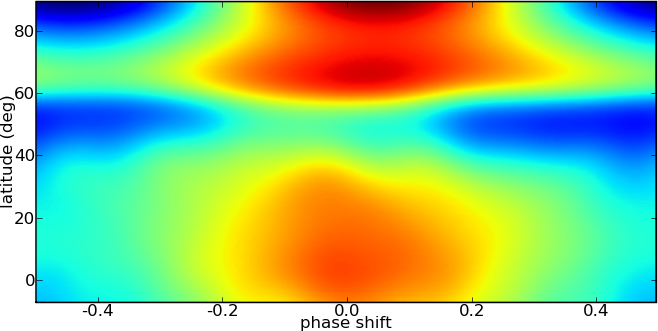}
\includegraphics[width=9cm]{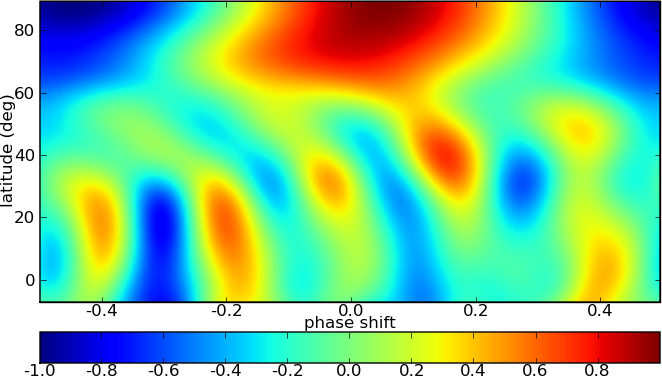}
\caption{Cross-correlation maps obtained from the comparison of the magnetic geometries reconstructed in July 2008 and September 2009. Results for each component of the magnetic vector are illustrated in a separate chart, with the radial, azimuthal and meridional components from top to bottom.}
\label{fig:cross}
\end{figure}

In spite of the very weak Zeeman signatures produced by the photosphere of Vega, the detection of a preferred rotation period is a good hint that the large number of spectra involved in the present study can partly compensate for the relatively high noise content of the Stokes V profiles and carry some useful, rotationally-modulated magnetic signal that can be modelled by means of Zeeman-Doppler Imaging.

The magnetic maps reconstructed for July 2008 and September 2009 are displayed in Fig. \ref{fig:mapjul08} and \ref{fig:mapsep09}, while the quality of data adjustment is illustrated in Fig. \ref{fig:zdi_comp}. The most recognizable structure on both maps is a spot of negative radial field centered on the visible pole, extending down to a latitude of 80 degrees and exhibiting a peak strength of about 5~G. This axisymmetric magnetic feature was anticipated by L09 from the shape of the phase-averaged Stokes V profile, where a negative Zeeman signature survives the averaging of observations corresponding to a large variety of rotational phases (showing that the magnetic feature producing the signature is approximtely axisymmetric), spanning a very limited range of Doppler velocities (i.e. concentrated over a small fraction of the stellar surface) and showing up very close to the line centre (as expected for a polar spot). 

Apart from this prominent polar magnetic structure, other reconstructed magnetic regions are less obvious to identify when comparing both epochs. Considering the radial field component in both years, a large positive patch is consistently visible around phase 0.7, while the opposite field polarity dominates around phase 0.25. Magnetic features consistently recovered at the one year interval are also visible in the azimuthal magnetic component, where a positive field region is visible around phase 0.6 in both maps. The comparison is even harder for the meridional field component, apart from the highest latitudes where positive field shows up between phase 0 and 0.5, while the opposite polarity is confined to rotational phases larger than 0.5.      

To help compare the two maps in a more objective and global way, we can estimate several numerical quantities reflecting some general properties of the surface field distribution and that can easily be derived from the spherical harmonic projection of the magnetic geometry. We first report a consistent average surface field level at both epochs, with 1.0~G in 2008 and 1.4~G in 2009. The field geometry is mostly poloidal, with $78\pm 2$\% and $58\pm7$\% of the magnetic energy reconstructed in the poloidal component in 2008 and 2009, respectively (the uncertainties are estimated through the reconstruction of several maps using different values of the input parameters, with $4^{\circ} \leq i \leq 12^{\circ}$ and $21$~\kms$\leq \vsin \leq 23$~\kms, see Petit et al. 2008). Considering the poloidal field component alone, the dipolar term hosts about $23\pm6$\% of the poloidal magnetic energy in 2008, versus $27\pm16$\% in 2009. We reconstruct just $10\pm3$\% (in 2008 and 2009) of the poloidal magnetic energy in the quadrupole, and $5\pm 1$\% in the octupole. We therefore find that a significant amount of the poloidal magnetic energy is reconstructed in modes with $\ell \ge 4$, with 62\% and 58\%, respectively. Finally, we estimate the percentage of energy showing up in the axisymmetric field component (taking the poloidal and toroidal components into account), concluding that the field is very far from axisymmetry, with only $21\pm 2$\% and $33\pm 2$\% of the energy stored in modes with $m = 0$.   

By replacing the rotation period of 0.7319~d deduced from our period search by the period of 0.525~d proposed by Aukdenberg et al. (2006), the reconstructed magnetic maps (not shown here) display a lower amount of magnetic energy in low-order poloidal components, with only 3\% and 6\% in the dipole (respectively for 2008 and 2009), 9\% and 7\% in the quadrupole, 5\% and 4\% in the octopole. The magnetic topology is also more axisymmetric, with 48\% and 28\% in modes with $m = 0$. Using the period of 0.663~d proposed by Hill et al. (2010), the main difference with the map derived using our best period value is a higher fraction of the poloidal magnetic energy reconstructed in the dipolar term, with 55\% and 84\% (associated to a lower fraction of the energy in the quadrupole and octopole). The field distribution is also less axisymmetric with this rotation period, with 15\% and 7\% of the energy in $m=0$. We note that, without any surprise, the polar magnetic region is consistently recovered using any of these rotation periods, since its axi-symmetric configuration does not produce any rotational modulation in its associated Zeeman signatures.

Another option to estimate the consistency of both maps, as well as a possible field evolution between the two epochs, consists in isolating strips of equal latitude in both maps to calculate their cross-correlation. The resulting cross-correlation maps obtained for each projection of the magnetic vector are plotted in Fig. \ref{fig:cross}. For the radial and azimuthal field components, a correlation parameter close to unity is achieved for a phase shift close to zero (at least if we ignore the intermediate-latitude domain of the azimuthal component), confirming that the maps computed from both independent data sets carry similar information (using the preferred rotation period proposed above). The correlation is also very good for the meridional field projection above latitude 60 degrees, but is much lower at low and intermediate latitudes (which is expected for a star with a low inclinaison angle, as stated in Sect. \ref{sect:procedure}). The absence of any significant phase shift between both epochs suggests that the surface differential rotation, if any, is probably very weak on Vega, so that the photospheric magnetic structure was not significantly distorted by a latitudinal shear within 1 year. 

\section{Discussion and conclusion}

Using 4 independent high-resolution spectropolarimetric data sets of Vega collected over more than one year with two different instruments, we confirm the detection of circularly-polarized signatures first reported by L09. The shape and amplitude of the signal show no significant differences between the successive observing runs, giving further support to a stellar origin of the signal. The lack of any detection in the diagnostic null spectrum, and the higher amplitude of the signature obtained while selecting photospheric spectral lines with a higher Land\'e factor, are strong evidences for a photospheric magnetic origin of this signal. 

Periodic variations of the signal are observed in our data and are consistent with a rotation period of $0.732 \pm 0.008$~d. This period is close to the value of $\approx 0.733$~d proposed by Takeda et al. (2008) from a careful modelling of individual spectral lines and the SED. This period is consistently recovered using two independent spectropolarimetric data sets, although this is not the only possible value if the different sets are considered individually. We do not find such repeated evidence in favor of the alternate value of $\approx 0.524$~d proposed by Aufdenberg et al. (2006) from interferometric observations. The data adjustment is also slightly degraded while using the value of $\approx 0.663$~d derived by Hill et al. (2010) from high resolution spectra. We infer that the stellar spin is close to a solid-body rotation, as suggested by the lack of evidence for any significant distortion of the magnetic field distribution over one year. New observations of Vega are obviously needed to further increase our temporal lever arm and better characterize the possible secular evolution of the magnetic field under weak large-scale motions (differential rotation or meridional flows).

The surface mapping of the magnetic field displays several striking differences while compared to magnetic geometries usually observed in magnetic intermediate-mass stars. Once averaged over the visible part of the stellar surface, the field strength does not exceed 1~G, while the observed large-scale fields of all Ap stars appear to be larger than a few hundred gauss at the magnetic poles (Auri\`ere et al. 2007). Furthermore, in spite of the weakness of the observed Zeeman signatures, we find repeated hints that the field structure of Vega is more complex than the large-scale fields of chemically peculiar stars (e.g. L\"uftinger et al. 2010), with about half of the poloidal magnetic energy reconstructed in spherical harmonics terms with $\ell > 3$ (except when using the rotation period of Hill et al. 2010). The complexity of this geometry is at odds with a fossil field hypothesis, since only low-order field geometries are expected to survive on long timescales. It could be argued that the young age of Vega (a few hundreds of Myr) could account for this complex field structure, but there is now growing evidence that strongly magnetic Herbig stars exhibit a very simple field topology (e.g. Alecian et al. 2008) whereas they are much younger than Vega (with typical ages of a few Myr for Herbig stars), confirming that the field topology we reconstruct is quite different from anything observed so far on intermediate-mass stars. 

A complex field structure would be more naturally expected in the presence of dynamo action. Several clues gathered from our observations could be reconciled with a stellar dynamo, starting from the fast rotation that is a critical ingredient to trigger large-scale dynamos. In the case of a dynamo (core-dynamo or radiative envelope dynamo), the observed high-latitude spot of radial magnetic field could possibly be interpreted as the surface emergence of flux tubes expelled through magnetic buoyancy, with a transit across the stably-stratified layers progressively deflected towards the spin axis by the action of the Coriolis force (a mechanism already proposed to account for the high-latitude emergence of cool spots on rapidly-rotating cool stars, Schuessler et al. 1996). 

Other information gained from the surface topology may help to distinguish between a core dynamo and a dynamo taking place in the radiative layers. We observe that, apart from the prominent polar magnetic region, a few other magnetic spots are reconstructed at lower latitudes (these low-latitude, non-axisymmetric magnetized regions are responsible for the rotational modulation of the Zeeman signatures). The observed surface location of these regions suggests they were emerging at the surface before experiencing an efficient polar deflection, implying that they may be formed in internal layers close to the photosphere, which tends to favor the radiative dynamo hypothesis against the core dynamo. 

Considering the hypothesis of a core dynamo, we note that the buoyant rise time from the core can become much longer than the age of Vega for weakly magnetized flux-tubes (Moss 2001, MacGregor \& Cassinelli 2003, MacDonald \& Mullan 2004, Mullan \& MacDonald 2005). The rise time is less problematic if taking place in the radiative envelope, as the magnetic field is generated higher in the stellar interior. Both the core and the radiative zone dynamo models involve a significant amount of differential rotation for the generation of a large-scale toroidal field, but provide few predictions about the expected surface flows. From our observations we can only argue that the global surface differential rotation seems to be very weak on the observable surface of the star, but the noise level in our data is too high to exclude the possible existence of photospheric zonal flows affecting only a fraction of the visible stellar hemisphere. 

An important clue to distinguish between the fossil or dynamo hypothesis would be to investigate the long-term stability of the observed magnetic geometry, as a dynamo-generated field is likely to experience some temporal variability. Without any detected changes in the magnetic geometry over a timespan of one year, we suggest that a long-term monitoring of the star is critically needed to probe further the temporal evolution of the field and decide between the various theories at our disposal. Accumulating more spectropolarimetric observations of Vega (and decreasing further the noise level) will also help to tackle more detailed studies of its magnetic structure. In particular, future work may include the study of Zeeman signatures associated to specific chemical species, in order to investigate the possible correlation between the magnetic field geometry and the surface abundance inhomogeneities, a phenomenon already well documented for Ap/Bp stars (e.g. Hubrig et al. 2006, Briquet et al. 2010).  

Finally, the high-\sn\ observation of a larger sample of normal intermediate-mass stars is of prime importance to establish the fraction of stars hosting a Vega-like magnetic field and determine the stellar parameters controlling the occurrence of such magnetism.     

\section*{Acknowledgements}

We thank the staffs of TBL and CFHT for their efficient help during these challenging observing runs. GAW acknowledges Discovery Grant support from the Natural Science and Engineering Research Council of Canada (NSERC) and from the Department of National Defence (Canada) Academic Research Program. VVG acknowledges grant support from the Centre National d'Etudes Spatiales (CNES, France). We are grateful to the referee, Dr Gautier Mathys, for comments that helped to clarify some aspects of this work.

\end{document}